%
\magnification 1150
\baselineskip 16 pt

%
%
 \def\ea{{\it et al\/}}
 \def\ie{{\it i.e.\ }}
 \def\Par{\par\vskip 7 pt}
 \def\eqn#1{ \eqno({#1}) \qquad }
 %
\font\bigggfnt=cmr10 scaled \magstep 3
 2
\font\bigfnt=cmr10 scaled \magstep 1

\leftskip .25 in
\rightskip .25 in
\noindent{\bigggfnt Limitations to the superposition principle:}\Par 
\noindent{\bigggfnt   Superselection rules in non-relativistic } \Par
\noindent{\bigggfnt       quantum mechanics} \Par
\vskip 12 pt

 \noindent {\bigfnt C Cisneros\dag, R P Mart\'{\i}nez-y-Romero\ddag, H N N\'u\~nez-Y\'epez$\|$\footnote\P{\noindent\rm On sabbatical leave from Departamento de F\'{\i}sica, UAM-Iztapalapa,\par \vskip -4 pt e-mail: nyhn@xanum.uam.mx},\par 
\noindent A\ L\ Salas-Brito\ddag \footnote\S{\noindent\rm 
Corresponding author.\par\vskip -4 pt 
On sabbatical leave from Laboratorio de Sistemas Din\'amicos, UAM-Azcapot\-zal\-co,\par \vskip -4 pt e-mail: salas@dirac.fciencias.unam.mx} } \Par
 \vskip 13 pt

\noindent \dag\  Laboratorio de Cuernavaca, Instituto de F\'{\i}sica, 
Universidad Nacional Aut\'onoma de M\'exico, Apartado Postal 139-B, C P 62190, Cuernavaca Mor, 
M\'exico\Par 

\noindent  \ddag\ Departamento F\'{\i}sica, Facultad de Ciencias, 
Universidad Nacional Aut\'onoma
de M\'exico, Apartado Postal 21-726, C P 04000, Coyoac\'an  D F,
M\'exico \Par

\noindent  $\|$ Instituto de F\'{\i}sica `Luis Rivera Terrazas', Benem\'erita Universidad 
Aut\'onoma de Pue\-bla, A\-par\-tado Postal J-48, C P 72570, Puebla  Pue, M\'exico \Par

\vskip 13 pt
 \centerline{\bigfnt Abstract}\Par

\noindent The superposition  principle is a very basic ingredient of quantum theory. 
What may come as a surprise to many students and even to some practitioners
of the quantum craft, is that  superposition  has limitations imposed by 
kinematical  or by dynamical requirements of the theory. The discussion of
such limitations is the main purpose of this article. For doing this we introduce
the central concept of classical superselection variables and of the superselection rules   
to which they are related. Some of their principal consequences are also discussed. 
The univalence, mass, and particle number superselection rules, all three resulting from 
kinematical requirements of  non-relativistic quantum mechanics, are next deduced. A brief 
discussion of dynamically induced superselection rules is next given and they are  illustrated 
with the simple example of the one-dimensional hydrogen atom.

\vskip 12 pt

\vfill
\eject

\centerline{\bigfnt Resumen}\Par

\noindent El principio de superposici\'on es un ingrediente important\'{\i}simo  de la mec\'anica 
cu\'antica; por ello, \'el que est\'e sujeto a ciertas limitaciones, que le son impuestas por 
consideraciones cinem\'aticas o din\'amicas, puede ser una sorpresa para muchos estudiantes y 
a\'un para algunos practicantes de las artes cu\'anticas. El prop\'osito de este art\'{\i}culo es 
la discusi\'on del or\'{\i}gen de tales limitaciones. Para ello introducimos la noci\'on 
fundamental de observable de superselecci\'on cl\'asico y de las reglas de superselecci\'on a las 
que aquellos dan siempre lugar.  Mencionamos y discutimos las principales consecuencias de la 
existencia de una 
regla de superselecci\'on. A continuaci\'on deducimos las principales reglas de superselecci\'on 
que ocurren en la mec\'anica cu\'antica no relativista y que tienen  un or\'{\i}gen puramente 
cinem\'atico, a saber, la de univalencia, la  asociada con la masa y la de n\'umero de 
part\'{\i}culas. Para terminar, discutimos en forma breve algunas reglas de superselecci\'on 
inducidas por la din\'amica e ilustramos todo ello con el ejemplo simple del \'atomo de 
hidr\'ogeno unidimensional. \Par
\bigskip

\noindent {\sl Classification numbers}: 03.65.Ca, 03.65.Bz \Par
\vfill
\eject

\noindent{\bf 1. Introduction}\Par

\noindent Linearity is one of the most basic properties of  quantum theory: 
any set of states of a quantum system, $|\alpha>$, where $\alpha$ is 
just a label for the quantum numbers needed to specify the 
state,  may be linearly superposed to obtain a new  state;  any combination
of the form

$$ |A>= \sum_\alpha a_\alpha |\alpha>,  \eqn{1}$$
 
\noindent where the $a_\alpha$  are complex numbers, stands for a new possible 
state of the system. The physical interpretation of each $a_\alpha$ is as the probability amplitude for the system to be found in precisely the component state $|\alpha>$. This requires the linearity of the mathematical structure of the theory, thus, all quantum operators are  
required to be linear. Linear combinations like (1) are usually referred to as {\sl coherent
superpositions} (Sakurai 1985). A brief discussion of coherent as contrasted to  incoherent 
superpositions is given in section 2. \Par

 If the set of states under consideration satisfy a quantum 
equation of motion (like Dirac's or Schr\"odinger's, for our argument it does not matter which 
one)

$$ H | \alpha> = i \hbar {\partial \over \partial t}|\alpha> , \eqn{2}$$

\noindent the new state (1), a coherent superposition of the original states, also 
necessarily does: 

$$ H |A>=  \sum_\alpha a_\alpha H|\alpha>= \sum_\alpha a_\alpha \, i \hbar\,
{\partial \over \partial t}|\alpha>= i \hbar{\partial \over \partial t}|A>. 
\eqn{3} $$

This is  the property which explained the puzzling ---in the early days of quantum theory--- 
phenomenom  of electron diffraction and that put it on the same footing as photon diffraction. 
The superposition principle is not only basic, it is also responsible, together with the 
interpretation of the $a_\alpha$ numbers as probability amplitudes, for some of the strange and 
conterintuitive aspects of quantum theory, like the so-called Schr\"odinger's cat paradox 
(Sakurai 1985, Sudbery 1986). \Par

 An  example of the superposition property of quantum systems can be 
illustrated considering the properties of spin-$1/2$ particles. The spin state of these particles 
can be found  pointing up $|+>$ or pointing down $|->$ along a previously specified $z$-direction 
---specified, for example, by the direction of a given  external field. These up and down states 
are independent from one another and, furthermore, any other spin states of the particle can be 
expressed as coherent superpositions in terms of these two; for example, spin states pointing in 
the $x$ or the $y$ directions can be respectively written

$$   |x>= {1\over \sqrt{2}}\left(|->\,+\,|+>\right), \qquad |y>= {1\over \sqrt{2}}
\left(e^{-{i\pi\over4}}|\,+>\,+\, e^{+{i\pi\over4}}|->\right). \eqn{4} $$

But, despite  being fundamental it is easy to realize that the 
superposition principle cannot hold unrestricted in every situation. 
Just  think of photons and electrons, then not 
mattering what we have just said about linearity and the 
superposition principle, no one has ever made sense of any  
superposition of electronic with photonic states (Sudbery 1986, 
Salas-Brito 1990). See the comment on supersymmetry at the end of section 2.1. \Par

 How can we understand this peculiarity within the 
general structure of quantum theory? The quantum justification  for such 
fact was found a long time ago (Wick \ea\ 1952) and involves the curious 
phenomenon referred to as a {\sl superselection rule}. To give a extremely 
brief description, a superselection rule exist whenever there are 
limitations to the superposition principle in quantum theory (Kaempfer 1965, Sudbery 1986), that is,
 whenever certain superpositions  cannot be physically realizable. In fact, the existence of superselection rules are needed to explain why we can treat certain observables like the mass or the charge of a particle as parameters rather than as full-fledged operators. \Par

It is somewhat peculiar that despite the additional insight into the subtleties of quantum theory that might be offered by superselection rules and of the fundamental role they have played in 
explaining puzzling phenomena in atomic, molecular and quantum field 
theories, the subject has not yet found a place in many of the excellent 
textbooks on quantum mechanics available. Perhaps this lack of attention is due to the initial belief that superselection rules were only important for quantum field theory (Streater and Wightman 1964, Lipkin 1973). In this article we want to contribute to alleviate this situation by first explaining how superselection rules fit in the general structure of quantum mechanics and then by offering some examples in the context of non-relativistic quantum mechanics. \Par

\noindent{\bf 2. Superselection rules in quantum theory}\Par

\noindent  {\sl 2.1 Superposing photons and electrons is forbidden}

\noindent Let us begin analysing the possibility of coherent superpositions of  photon and electron states. Any such superposition could, at least in principle, be written in the form

$$ |\Psi>=  a_p\,|\hbox{photon}> + \, a_e \,|\hbox{electron}>.   \eqn{5}$$

\noindent where some irrelevant details ---irrelevant at least for the present 
discussion--- are left out of the description of the states.   Before 
discussing (5), recall that whereas electrons have a half integer spin,
$s_{{}_{\hbox{\sevenrm electron}}}=1/2$, a property which makes them  fermionic particles, photons have 
an integer spin, $s_{{}_{\hbox{\sevenrm photon}}}=1$, which makes them  bosonic particles. 
Bosons and fermions differ in many ways. Fermions comply 
with Pauli's exclusion principle, they cannot `stand' another identical particle in the same state. Bosons, on the contrary, `like' to crowd one over another in the same quantum state (Feynman 1965). Another crucial difference between fermions and bosons is found in their behaviour 
under rotations. Consider a bosonic state and rotate the system by $2\pi$, a rotation which is obviously expected to let the physical content of this and any other state unchanged, the state does indeed not change. On the other hand, a fermionic state has a different behaviour, under 
that same rotation it changes its phase by $ \pi$, \ie\ the state ends multiplied times $-1$. Even so, there is nothing to worry about since the quantum states are  determined up to a normalizing factor ---or, to describe the situation in other words, the important thing is the modulus of the state since it is related with the actual probability and not just 
with its amplitude--- and thus the final state is also physically indistinguishable from the initial one. However, if we rotate by $2\pi$ not a fermionic or a bosonic state separately but the superposition $|\Psi>$, the rotated state is  then

\vskip -5 pt

$$ {\cal R}(2\pi) |\Psi> = |\Psi'>= a_p|\hbox{photon}> - \, a_e 
\,|\hbox{electron}>. \eqn{6} $$

\noindent  Given that $<\Psi|\Psi> \neq <\Psi'|\Psi'> $, under the rotation the superposition  becomes a state essentially different from the original one. This implies that a  superposition of the form  (5) is  devoid of physical meaning unless, of 
course,  $a_p=0$ or $a_e=0$, or, in other words, only if a coherent superposition 
is not allowed (Lipkin 1973). Thus, to be consistent with the  kinematical 
requirements of rotations, quantum theory is forced to restrict the 
applicability of the superposition principle and to impose a superposition 
rule  between photons and electrons or, to state the restriction in completely general terms, to impose a superselection rule forbidding superpositions of any fermionic with any 
bosonic states whatsoever ---see Weinberg (1995) for a recent discussion of the generality and applicability of this restriction. This superselection rule is usually called {\sl univalence} 
(Wick \ea\ 1952, Streater and Wightman 1964, Hegerfeld \ea\ 1968). \Par 
 
To avoid possible misunderstandings, it is important to pinpoint that there 
is a way of making sense of a linear combination like (5) obviously not as a coherent 
superposition which we have just proved to be impossible but  as one of the so-called mixed 
states or incoherent superpositions. That is, an expression like (5) can have meaning only for 
describing a {\sl statistical ensemble} with a fractional population $a_p$ of photons and a 
fractional population $a_e$ of electrons (Landau and Lifshitz 1977, Sakurai 1985, 
N\'u\~nez-Y\'epez \ea\ 1988, Salas-Brito \ea\ 1990) ---as, for example, could be necessary to 
describe the situation in the interior of a switched-on floodlight. \Par

We should mention also the case of supersymmetric quantum mechanics (Robinett 1997). In 
supersymmetric systems, for each 
bosonic state there is an associated fermionic partner required by the formalism. 
 The supersymmetry allows the existence of a sort of coherent  superposition of 
fermionic and bosonic states.  That is, we can get bosonic and
 fermionic states in the same representation multiplet but, as a matter of
 fact, this behaviour should not be considered a restriction to the
 univalence superselecction rule just discussed. The key to understand the point is the existence 
of anticommuting Grassmann numbers in the supersymmetric formalism which, under the rotation 
operator ${\cal R}(2\pi)$, change again the sign in Eq (6) and finally recover the plus sign 
in the rotated system thus behaving as if it were a pure bosonic state (Cooper \ea\ 1995). \Par


\noindent{\sl 2.2 Coherent and incoherent superpositions}\Par

\noindent Before procceding any further, it is convenient to review the difference usually made in quantum mechanics between coherent and incoherent superpositions .  Take, for example, the spin-$1/2$ particle states mentioned  in the introduction. The most general spin state for one of those particles  can be expressed in the form of a coherent superposition

$$ |\phi> = a_+ |+> + a_- |->  \eqn{7}  $$

\noindent where the components $a_+$ and $a_-$ are, in general, complex numbers which, if the state is assumed normalized to 1, comply with $|a_+|^2 + |a_-|^2=1$. The superposition (7) always describes a state with a well-defined spin. States  arising as coherent superpositions are sometimes termed pure states. The quantum expectation value of any observable $O$ with respect to the pure state $|\phi>$ can then be expressed as

$$ \eqalign{
 <\phi| O|\phi>= & |a_+|^2 <+|O|+> + |a_-|^2 <-|O|-> \,+\cr 
   & a_+^*a_- <+|O|-> + a_-^*a_+<-|O|+>; }\eqn{8} $$

\noindent the last two terms of equation (8) are called  interference terms. These interference terms,  basically arising from the information about the relative phase of the states $|+>$ and $|->$ given by the complex numbers $a_+$ and $a_-$,  are an  unavoidable characteristic of pure states. \Par

However, states of the form (7) are not general enough to describe every possible spin situation. Sometimes  the need to describe a so-called {\sl mixed} spin state, or, as it is also called, an {\sl incoherent superposition} of spin states, might arise. For example, if the spin of some  {\sl unfiltered beam} of particles is required, a mixed state is necessary to describe the random spin orientations in the beam. It is not possible to do this  using coherent superpositions, like in (7), which are only capable of describing particles whose spin is pointing in some definite direction. \Par

To deal with the random spin situation,  mixed states of the form 

$$ w_+ |+> + w_- |-> \eqn{9} $$ 

\noindent  must be introduced. In the situation described by (9), the numbers $w_+$ and $w_-$ are {\sl necessarily  real} and should be interpreted as just statistical weights  describing the fractional  population of the state, thence, the usual condition $w_+ + w_-=1$ imposed on them; they should {\sl never} be interpreted as components in the two-dimensional spin space. Notice thus the complete lack of information on the relative phase between the up and the down states implied by the reality of $w_+$ and $w_-$. In particular this means that interference terms  cannot occur if incoherent superpositions are used. In the case of the mixed state (9), the measurement of any property $O$  is calculated using not the quantum mechanical expectation value  but by taking the {\sl ensemble average} of $O$ (Sakurai 1985) between the mixed state (9) 

$$ [O]\equiv w_+ <+|O|+> + w_- <-|O|->.  \eqn{10} $$

\noindent This expression lacks the interference terms  of the quantum expectation value (8) calculated using   pure states. We can say that incoherence implies the absence of interference terms. As the next section shows,  in systems with superselection rules there necessarily are states which exist only as mixed, incoherent superpositions and, therefore, that can never interfere whith one another.\Par


\noindent  {\sl 2.3 Superposition rules in the formalism of quantum theory}\Par

\noindent In this section we explain how superselection rules are embedded into the
structure of quantum theory and discuss some of their consequences. \Par

Let us begin with a definition: If in a  quantum  system there exist a hermitian operator $G$ 
which commutes with all the observables $O$, that is, such that

$$ [G, O] = 0, \qquad \hbox{for every observable } O,   \eqn{11}$$

\noindent and if $G$  is  found {\sl not} to be a multiple of the identity operator, then, it is 
said that a superselection rule induced by $G$ is operating in  the system. This is in fact 
closely related to the failure of the superposition principle as is shown in the next 
subsection.\Par

 Since $G$ commutes with every operator,  it commutes, in particular, with the Hamiltonian and 
then, it necessarily corresponds to a conserved quantity. We must distinguish this property of 
$G$ from the corresponding one of ordinary conserved quantities, like the energy $E$, since 
physically realizable states exist that are not eigenstates of the Hamiltonian. For example, 
since we know a superselection rule does {\sl not} act, nothing prevents us from considering an 
arbitrary nonstationary state  $|\psi>=|n, l, m_s> + \exp(i \delta) |n', l', m_s'>$  (where $n$, 
$n'$, $l$, $l'$, $m_s$  and $m_s'$  are standard hydrogen-like quantum numbers and $\delta$ an 
arbitrary real number) of an hydrogen atom as physically realizable. On the other hand, when a 
superselection rule induced by a certain operator $G$ acts on a system, {\sl every} physically 
realizable state  must be an eigenvalue of $G$ and not every possible superposition can represent 
a physical state. \Par

 The property of commuting with every other operator makes the variable $G$ sharply measurable in 
every situation being thus rather similar to a $c$-number. It is not surprising then that any 
such operator is  termed by some authors a classical observable of the quantum system. Despite this,
we must exercise some care in the use of the term since it is used with a very different meaning 
in other quantum contexts; to avoid conflicts or misunderstandings, we here have decided to  use 
the term {\sl classical superselection variable} (CSV for short) for these operators. Thus we can 
say that every CSV has an associated superselection rule, these two notions are really two 
manifestations of the same phenomenom. \Par

CSV's in generic nonrelativistic  quantum systems are the mass and 
the electric charge (Bargman 1954, Foldy 1954); in molecular physics they can be the chirality of 
a molecule or its tertiary structure (Primas 1981, Pfeifer 1981, 1983, M\"uller-Herold 1985); in 
many particle systems, it is  the particle number or the total mass (Kaempffer 1965).  \Par

Superselection rules and classical superselection observables can even occur in rather simple 
systems. For example, in the one-dimensional hydrogen atom with interaction $V(x)=-e^2/|x|$; the 
CSV is the one-dimensional Laplace-Runge-Lenz vector, that is, it is just the 
side of the singularity in the potential in which the particle is confined to move. This was shown 
by Boya \ea\ (1988) in an interesting and easily
readable article. The superselection rule forbidding superpositions among such confined states  
was shown to exist  by N\'u\~nez-Y\'epez \ea\ (1988, 1989), see in particular a contribution to
this journal where the problem is discussed from an elementary standpoint (N\'u\~nez-Y\'epez \ea\ 
1987). \Par

Let us denote the eigenstates and eigenvalues of $G$ as

$$ G |g_m; \alpha_m> = g_m | g_m; \alpha_m>,  \eqn{12} $$

\noindent where $\alpha_m$ stand for every other quantum number 
needed to specify the eigenstates. As every operator commutes 
with $G$ its eigenstates can always be chosen as eigenstates of the other 
dynamical variables of the system. Given this, any state $|S>$ in the Hilbert 
space of the system can be always expressed as a linear superposition 
of the eigenstates of $G$

$$ |S> =\sum_m a_m |g_m; \alpha_m>;   \eqn{13}  $$

\noindent thus, the complete Hilbert space  of the system ${\cal H}$, is 
naturally decomposed into a collection of disjoint and mutually orthogonal  subspaces 
${\cal H}_g$ (this follows since they correspond to different quantum numbers of a Hermitian operator, and, therefore, $<~g_s; \alpha_s|g_m; \alpha_m>=0\quad (s\neq m)$), each identified by a 
different eigenvalue of $G$. The ${\cal H}_g$ subspaces, are called the 
{\sl coherent subspaces} or {\sl coherent sectors} induced by the 
superselection rule.  Recalling the definition of a direct sum between vector spaces, as the 
existence of a unique expansion like (13) for every element in the original space and where every 
term in the expansion belongs to only one subspace (Shilov 1977), it is easy  to see that the 
direct sum of all the coherent subspaces

$$ {\cal H}=  {\cal H}_1\oplus{\cal H}_2 \oplus\dots,  \eqn{14}  $$

\noindent with as many terms as different eigenvalues of $G$ exist, is the complete Hilbert space 
of the system. \Par

One must be somewhat cautious with the statements (13) and (14), 
because, although the expansions are allowed in principle, the very existence of  the superselection rule forbids any {\sl physically meaningful} superposition of states in different 
coherent subspaces. That is, a superselection rule forbids superposing 
states with different values of the eigenvalue $g_m$, as is  explicitly
shown below in subsection 2.4. An additional consequence of the existence of a superselection 
rule is that not every Hermitian operator, even if related to a supposedly 
good physical variable, may represent an observable. For this to be true, 
the operator should not induce transitions between the different coherent 
sectors of the system. This property is also proved below in subsection 2.5.
\Par

\noindent  {\sl 2.4 Impossibility of superposing states belonging to different coherent 
sectors}\Par

\noindent To show that any superposition of eigenstates of $G$ with 
different eigenvalues is not physically admissible, that is, that it cannot be observed in nature,
consider first that, on the contrary, the state

$$ |u>=\sum_m u_m |g_m; \alpha_m> \eqn{15}  $$

\noindent is physically reasonable. Since $G$ commutes with everything, the state 
(15) should be an eigenstate of every operator in some complete set of 
commuting observables of the system. Denote this set as 
$O_s, \quad s=1\dots, N$, thus

$$ O_s |u> =  o_s |u>. \eqn{16} $$ 

\noindent As a consequence  of the superselection rule, the state
 $|g_m; \alpha_m>$ is also an eigenstate of $G$ with eigenvalue 
$o_s^m$; then, another way of express state (16), is
as

$$ O_s |u> = \sum_m u_m o_s^m |g_m; \alpha_m>,  \eqn{17} $$

\noindent but,  all the $|g_m; \alpha_m>$ being linearly independent from one 
another, equations (15), (16) and (17) imply that $o_s^m=o_s$ for 
every $m$-value.  Next consider the modified  state
 
 $$ |u'> = \sum_m u_m \exp(i \delta_m) |g_m; \alpha_m>, \eqn{18} $$

\noindent which is just (15) with the phases of each component arbitrarily 
modified, and apply the operator $O_s$ to it. In this way the result

$$ O_s|u'>= o_s \sum_m u_m \exp(i \delta_m)  |g_m; \alpha_m>= o_s |u'> \eqn{19}  $$

\noindent is easily obtained. On comparing (19) with (16), we can see that both states, 
$|u>$ and $|u'>$, have precisely the same quantum numbers. Therefore, the two 
states (16) and (19) are completely indistinguishable from one another. No physical 
process could be able to determine the relative phase between $|u>$ and $|u'>$. But they are 
also completely different vectors in Hilbert space; the only way of avoiding
 a contradiction is by altogether forbidding superposition of states 
in different coherent subspaces. This proves the statement and establishes the connection between 
the existence of operators commuting with everything and the existence of limitations to the 
superposition principle. \Par

Notice also that, as we have already said, the only way of give physical meaning to superpositions
 like (15) or (18) is by interpreting them as mixed or impure states, never as coherent 
superpositions. Thus, in a system with  superselection rules, the superposition principle only 
holds  unrestricted within each coherent subspace, superpositions of states from different 
coherent subspaces can never describe physical states. 
\Par

\noindent  {\sl 2.5 Nonexistence of transitions between different sectors}\Par

\noindent It is very easy to show the vanishing of any matrix element of an 
observable $O$ between states belonging to different coherent sectors 
${\cal H}_n$ and ${\cal H}_m$, $m\neq n$. To acomplish this, keep in mind 
that the Hermitian operator $G$ commutes with everything else and that its eigestates can be therefore chosen to have common eigenvalues with any operator, thus

$$  
 <g_m; \alpha_m| O  |g_n; \alpha_n> =o <g_m; \alpha_m | g_n; \alpha_n>;
\eqn{20} $$   
 
\noindent but any states with different quantum numbers  (\ie\ such that $n\neq m$) are necessarily 
orthogonal, then

$$  
 <g_m; \alpha_m | g_n; \alpha_n>=0,\qquad m\neq m.\eqn{21} $$    

\noindent This proves that no observable can ever induce transitions between
 states in different coherent sectors of the Hilbert space. This means, for 
example, that, according to the univalence superselection rule, it is impossible  to produce a single electron from a single photon in any conceivable physical process. \Par

\noindent  {\sl 2.6 The most general superselection rules}\Par

\noindent As it must be clear by now,  superselection  rules are very important
properties of quantum systems; any quantum theory must be formulated 
taking them into account. The most general superselection rules that, 
it is believed, must hold in any quantum theory ---we are talking here, just in this subsection,
 of the most general relativistic quantum theories--- are associated first with the existence of fermions and bosons, \ie\ the univalence superselection rule, and then  to every absolutely conserved quantum number: there are superselection rules associated  to the electric charge, to the barion number, and to the three lepton numbers (Sudbery 1986); this means that all physical states of the basic building blocks of out universe should, according to this point of view, be sharp eigenstates of the operators associated with those properties. But even within the realm of non-relativistic quantum mechanics there are very interesting  superselection rules, which, besides, can be obtained by so elementary calculations that they could be part of almost any intermediate course in quantum mechanics. This is exhibited in the next section. \Par


\noindent {\bf 3. Superselection rules in non-relativistic quantum 
mechanics}\Par

\noindent The superselection rules acting in non-relativistic quantum mechanics and
the CSV associated with them are discussed in this 
section. We discuss the  univalence superselection rule separating the bosonic from 
the fermionic sectors in the Hilbert space of nonrelativistic quantum mechanics, the 
superselection rule separating the different mass sectors, which allow the mass to be treated as 
just a parameter in the Schr\"odinger equation, and the superselection rule separating states 
of different particle numbers in Fock space. The electric charge superselection rule also 
operates in nonrelativistic quantum mechanics but, as its derivation involves field operators  
and the equations of electrodynamics (Aharonov and Susskind 1667, Strocchi and Wightman 1974), 
the derivation of this rule is outside of the scope of the article.\Par

\noindent  {\sl 3.1 Univalence superselection rule}\Par

\noindent In section 2.2, following the arguments of Wick \ea\ (1952), we have 
already sketched a proof of this superselection rule valid for any rotationally invariant theory. In this section, following the excelent discussion of M\"uller-Herold (1985), we rederive the univalence superselection rule using arguments closer to the spirit of non-relativistic quantum mechanics. We just have to invoke some consequences of the existence of indistiguishable particles. \Par

  Think of a system of  any two particles (not necessarily assumed identical at this moment) with equal spin ---for the sake of simplicity we describe the argument using just two-particle states but it can be generalized to states of any number of particles (Galindo \ea\ 1962). We can represent the state by the ket $| 1, 2>$, where the content of the first slot labels the spatial and spin degrees of freedom of the first particle and the content of the second slot labels the same properties of the second one. \Par

  If $P$ stands for the  permutation operator for two particle states, we must have 

$$  P| 1, 2>= | 2, 1>.  \eqn{22} $$

\noindent  The only eigenvalues of $P$ are $+1$ and $-1$, as  it is very easy to prove by applying $P$ twice to any of its eigenstates. The eigenstates of $P$ are the symmetric and antisymmetric states under interchange of particles; these states are called, respectively, bosonic and  fermionic states (Feynman 1965). Moreover, every state vector in Hilbert space can be 
classified as fermionic or bosonic depending only on the spin of the system. This fundamental result, known as the spin-statistics theorem, has a rather difficult  proof. The only attempt we know towards  simplifying the  proof of the spin-statistics theorem is in Feynman's 1986 Dirac memorial lecture (Feynman and Weinberg 1987).   \Par

If the particles being described by (22) happen to be identical in the quantum sense, that is if they are, for example, two electrons or two protons, no property of the system can depend on the order in which their labels appear in (22) (Lipkin 1973). That is, all matrix elements of any observable $O$, between two identical-particle states $|1, 2>$, should be invariant under permutations 

$$  
<1,2|O|1,2>= <2,1|O|2,1>= <1,2|P^\dagger O  P|1,2>,
         \eqn{23} $$

\noindent since, being the particles absolutely indistinguishable, no physical property could depend on the order in which we had decided to write their labels. Therefore, as follows from (23) and the unitarity of the permutation operator (Sakurai 1985), every observable of the system should commute with $P$, \ie\ the permutation operator is a CSV and thus induces a superselection rule. Now, consider the projection operators

$$ P_{\pm} = {1\over 2} (1\pm P).   \eqn{24}  $$

\noindent If $|\phi>$ is an arbitrary state, then $P_+|\phi>$ is a 
bosonic and $P_-|\phi>$ is a fermionic state. These projection 
operators split the Hilbert space ${\cal H}$ in bosonic 
${\cal H}_+=P_+ {\cal H}$ and fermionic ${\cal H}_- = P_- {\cal H}$
 sectors such that ${\cal H}= {\cal H}_+ \oplus {\cal H}_-$. \Par

If the state vectors $|\phi_\pm>$ belong to ${\cal H}_\pm$ and $O$ is any 
observable, then its matrix elements between any fermionic and any bosonic states are 

$$ \eqalign{ < \phi_-| O| \phi_+> =&< \phi_-| OP| \phi_+>\cr
                          =&< \phi_-|P O| \phi_+>\cr
                          =&< \phi_-|P^\dagger O| \phi_+>\cr 
                          =&-< \phi_-| O| \phi_+>\cr  =&0. }\eqn{25} $$  
                              
\noindent since, in the order in which they are used in (25), $P|\phi_+>= + |\phi_+>$, $P$ commutes with any $O$,  $P$ is an unitary operator and $P|\phi_->= -|\phi_->$. Thus, no observable can ever induce transitions from bosonic to fermionic states or viceversa.  Furthermore, by applying $P$ to any linear combination of a fermionic with a bosonic state,

$$P( a_+ |\phi_+> + a_-|\phi_-> )=  a_+ |\phi_+> - a_-|\phi_->, \eqn{26} $$

\noindent we get a state linearly independent from the original one. 
However, as this state ought to  indistiguishable from the original one, the only possible conclusion is that a superselection rule is in operation. The operator $P$ is the CSV inducing the univalence superselection rule which, accordingly, has to be regarded as a direct consequence of the existence of indistinguishable particles. \Par

\noindent  {\sl 3.2   Superselection rule separating different mass sectors}\Par

\noindent In every elementary application of non-relativistic quantum mechanics, mass is treated 
as a aprameter. For example, in dealing with the Schr\"odinger 
equation, it is always implicitly assumed to be a $c$-number. Moreover, 
states superposing different masses are conspicuously absent of every 
treatment. Mass seems to comply with a superselection rule. That this 
statement is true follows from the Galilean  invariance of non-relativistic
quantum mechanics as was originally noticed by Bargman (1954). \Par

Galilean transformations in quantum mechanics are represented by unitary 
operators $U_G$ acting upon the states of a system. The change in a state functions when both a displacement of the origin by the constant quantity ${\bf r}$, and a change between two reference frames moving respect one another at a constant velocity ${\bf v}$, occur simultaneously; that is, when  the  Galilean transformation ${\bf q}'={\bf q} + {\bf v}t + {\bf r}$ is performed, can be  represented by the  operator (Kaempffer 1965)

 $$U_{G}=\exp(im{\bf v}\cdot {\bf q}' -i m v^2 t/2). \eqn{27}$$  

The action of $U_{ G}$ on a quantum state $|\Psi>$ describing a particle with  mass $m$,  is 

$$ |\Psi'>=U_{ G} |\Psi > = \exp\left( i m ({\bf v}\cdot{\bf q}'- 
   v^2/2)\right)|\Psi>; \eqn{28} $$

\noindent  therefore $<\Psi'|\Psi'>=<\Psi|\Psi>$ and Galilean invariance is manifest in 
nonrelativistic quantum mechanics.  Selecting ${\bf r}=0$, specializes (27)  to the unitary 
operator associated to a pure Galileo transformation: $U_{\bf v}$; whereas selecting  ${\bf v}=0$, specializes (27) to the operator associated to a pure spatial displacement: $U_{\bf r}$. \Par

  It must be clear that the  combined action on $|\Psi>$ of a spatial displacement $U_{\bf r}$, a 
pure Galilean transformation $U_{\bf v}$, the opposite displacement  $U_{\bf -r}$ and the inverse 
Galilean transformation $U_{\bf -v}$, just amounts  to an identity transformation. Its  net 
effect, as you may easily check,  is again just a change in the phase of the state:

$$ U_{I} |\Psi > = \exp(im{\bf v\cdot r})|\Psi >, \eqn{29} $$

\noindent where we  defined $U_{I}\equiv U_{\bf -v}U_{\bf -r}U_{\bf v}U_{\bf r}$.
The important point in (24) is that the phase of the state changes depending  on the mass of 
the particle. The transformation $U_I$ is thus seen to be of trivial consequences as long as we 
are dealing with definite mass states. This change of phase becomes crucial, however, if we 
consider  states $ |\phi_1> $ and $ |\phi_2>$, describinging particles with different masses 
$m_1$ and $m_2$. Since if we  try any superposition like  $|\phi> = a_1|\phi_1> + a_2|\phi_2>$, 
the new state $|\phi>$  transforms under the identity-equivalent transformation $U_I$ as
\vskip - 6pt
$$    |\phi'>\equiv U_{I} |\phi> =a_1\exp(im_1{\bf r\cdot v})|m_1> + a_2
\exp(im_2{\bf r\cdot v})|m_2>. \eqn{30} $$

\noindent From (30), you can  see that, if $m_1\neq m_2$ and if ${\bf r}\neq 0$ or 
${\bf v}\neq 0$, the states $|\phi>$ and  $|\phi'>$ become essentially different.  Then, again to 
avoid inconsistencies, we have to conclude that mass induces a superselection rule in 
nonrelativistic quantum mechanics.   Mass is  indeed a CSV in nonrelativistic 
quantum mechanics, it   can always be sharply measured in any situation---but, notice that this 
is a direct consequence of Galilean invariance,  in relativistic quantum  theory mass cannot  be 
always sharply measured  and there is not a corresponding superselection rule.  \Par

 The  whole Hilbert space associated with nonrelativistic quantum mechanics hence decomposes in a 
direct sum of sectors corresponding to each possible value the mass can take. This makes 
nonrelativistic quantum mechanics utterly inadequate for describing states endowed with a mass 
spectrum or for describing transformations between particles. In ordinary quantum mechanics no 
operator can exist which produces transitions between states characterized by different masses. 
\Par

\noindent {\sl 3.3 Particle number superselection rule}\Par

\noindent Given Bargman's mass superselection rule, it is rather easy to obtain
another,   the closely related superselection rule separating states with different particle 
numbers in Fock space. This superselection rule must be rather intuitive because nobody seems to 
care about superposing, say, 3-particle states with 11-particle states, it always sounds more 
reasonable to deal with 14-particle states from the start.\Par

The proof of this rule follows from the following argument (M\"uller-Herold 1985), given a 
 state $|n>$ describing $n$ identical particles, each with the same mass $m$, this state should 
transform, under the Galilean transformation $U(I)$---which you surely recall, is equivalent to 
an identity transformation--- of section 2.2 as

$$ U(I)|n>= \exp(i n m{\bf v\cdot r})|n>.   \eqn{31}   $$

\noindent From this result and from the fact that the states of $n$ 
particles have a total  mass $n m$ different, in general, from the mass $n' m$ of 
 $n'$-particle states, the argument of the previous section  unavoidably leads  to the  
conclusion that any  superposition of $n$-particle with $n'$-particles states must describe a 
mixed state as long as  $n \neq n'$. Therefore, another CSV of nonrelativistic 
quantum mechanics is the particle number operator $N$. \Par

\noindent {\bf 4. Dynamically induced superselection rules}\Par

\noindent In the previous section we have introduced some of the superselection rules 
acting in non-relativistic quantum theory. The common point of all of them
is their kinematical origin. They mainly come about because certain 
transformation properties  must hold for every possible quantum state and to 
avoid inconsistencies within the theory the superposition of certain states 
must be avoided. However, not every superselection rule is kinematical in 
origin there are some that occur when interactions are turned on. Examples 
of these are the superselection rules induced by the many body couplings
in macroscopic bodies. Manifestations of these superselection rules are
the CSV's like the temperature or the chemical potential, 
that appear as the thermodynamic limit is approached (Primas 1981, Pfeifer 1981, 
Zureck 1982, Putterman 1983). \Par

\noindent {\sl 4.1 Optical isomers in molecular physics}\Par

 \noindent Another  interesting example of dynamically induced superselection rules, is the  solution to the so-called paradox of optical isomers proposed in Pfeifer's dissertation (1980). This superselection rule is basically produced by the coupling of molecular degrees of freedom to the   electromagnetic field surrounding the molecule. Under certain conditions this coupling generates the superselection rule that explains, for example, why  certain chemical compounds like serine 
or even a household substance like sugar, can be produced in either right-handed or left-handed 
forms, but never in a non-chiral pure ---in the quantum not the chemical sense--- form 
(M\"uller-Herold 1985). The only possibility, if we do not want them in their levogirous nor in 
their dextrogirous forms,  is producing them in a mixed quantum state a so-called racemic mixture.
 The CSV that has been found associated to this superselection rule is  chirality 
(Pfeifer 1980, 1981).\Par

   Most  superselection rules with a dynamical origin have in common a certain complexity in 
their causes, since the explanation usually involves systems with an infinite number of degrees 
of freedom, coming about from either the need of taking  the thermodynamic limit as in a 
macroscopic system, or by  the existence of couplings to  external fields as in Pfeifer's rule. 
It is very interesting, thus, that a rather simple one-dimensional system is capable of showing a 
dynamically induced superselection rule: the one-dimensional hydrogen atom. \Par

\noindent{\sl 4.2 Superselection rule in the one-dimensional hydrogen atom}\Par

\noindent The Hamiltonian of the one-dimensional hydrogen atom  is

$$  H= {p^2\over 2m}- {e^2\over |x|}.      \eqn{32}$$

\noindent where $m$ is the mass and $e$ electric charge of an electron. The Hamiltonian (32) is 
parity invariant: the coordinate inversion $x\to -x$ leaves (32) unchanged. The normalized 
eigenstates of the system, in atomic units $\hbar=m=e=1$, and in the coordinate 
representation, can be written (N\'u\~nez-Y\'epez \ea\ 1987) as right 

$$ \psi^n_+(x)= <x|+n>=\cases{\quad
     0     & if $x\leq 0$,\cr
     \left( {4\over  n^5(n!)^2} \right )^{1/2} (-1)^{n-1}xL^1_{n-1}(2x/n)\exp(-x/n)& if $x>0$,\cr
 } \eqn{33a}$$

\noindent and left
\vskip -5 pt
$$ \psi^n_-(x)= <x|-n>=\cases{
          \left( {4\over  n^5(n!)^2} \right )^{1/2} (-1)^{n}xL^1_{n-1}(-2x/n)\exp(+x/n)& if $ x < 0$,\cr
\quad 0 & if $x\geq0$,
 } \eqn{33b}$$

\noindent  eigenstates, where the $L^1_{n-1}(z)$ are generalized Laguerre polynomials and $n$ is a 
strictly positive integer---please note that the eigenfunctions in (N\'u\~nez-Y\'epez \ea\ 1987)  
have certain misprints which have been corrected in (33). The corresponding  energy eigenvalues are exactly the same as in the three-dimensional case $E_n=-1/2n^2$. As it is rather clear in (33), the {\sl attractive} potential energy term of the system acts like an impenetrable barrier requiring $\psi^m_{\pm}(0)=0$ for all $m$-values ---if this condition is not met then the 
Hamiltonian fails to be Hermitian. Such properties confine ---and  explain the names given to the 
states (33)--- the particle to move either to the right or to the left of the origin without any 
chance of escaping from one region to the other. But,  as the one-dimensional hydrogen atom is 
reflection invariant, we can try to match the eigenfunctions (33) to form parity eigenstates

$$ \eqalign{
          \phi^n_{\hbox{e}}(x)=&  {1\over \sqrt{2}}(\psi_+^n(x) +  \psi_-^n(x)), \cr
          \phi^n_{\hbox{o}}(x)=&  {1\over \sqrt{2}}(\psi_+^n(x) -  \psi_-^n(x)),}
 \eqn{34}  $$

\noindent and, in this way, extend the states to the whole $x$-axis.  The problem is that for the 
even $\phi^n_{\hbox{e}}$ and the odd $\phi^n_{\hbox{o}}$ states to be realizable states of the 
system they must be independent from each other.  It is easy to see, however, that the Wronskian 
determinant $W(\phi^n_{\hbox{e}}, \phi^n_{\hbox{o}})$ always vanishes and thus that the definite 
parity states (34) cannot represent independent states of the one-dimensional hydrogen atom. This 
means that the relative phase of the states $\psi^n_+$ and $\psi^n_-$ is intrinsically irrelevant 
in any supposed coherent superposition. Any form of superposing $\psi^n_+$ with $\psi^n_-$ is  
thus devoid of physical meaning. We must conclude that a superselection rule operates in the 
system (N\'u\~nez-Y\'epez \ea\ 1988). \Par

The CSV $G$ in the one-dimensional hydrogen atom is the side of the singularity in
 which the particle is confined to move, or, to be more formal, is the one-dimensional 
Laplace-Runge-Lenz vector. Notice that $G$ is here also a  sort of chiral property as is the 
case with the CSV in Pfeifer's superselection rule. Parity, not commuting with 
$G$, becomes unobservable; we can say that it is spontaneously broken in the one-dimensional 
hydrogen atom. These properties have been established from a purely classical point of view in 
(Boya \ea\ 1988) and from a quantum point of view in (N\'u\~nez-Y\'epez \ea\  1989, 
Mart\'{\i}nez-y-Romero \ea\ 1989a,b). In a way of speaking, the superselection rule  in the 
one-dimensional hydrogen atom is an extremely simplified `bare-bones' version of that of 
Pfeifer. \Par

We think that this example suffices to exhibit some of the differences  between the 
generally applicable kinematical superselection rules of the previous sections and the 
superselection rules required by specific features of the interaction, as is the case of the 
one-dimensional hydrogen atom. \Par

\noindent {\bf 5. Summary}\Par

\noindent The essential concepts of classical superposition observables and superselection rules 
have been introduced and 
its principal consequences have been discussed within the framework of nonrelativistic quantum 
mechanics. We have exhibited how certain kinematical requirements are capable of constraining the 
applicability of the superposition principle. Using the basic machinery of nonrelativistic 
quantum mechanics we have derived three of the basic superselection rules that operate in that 
theory, namely, univalence, mass and particle number.  It is surprising that these very 
interesting and important ideas are almost never discussed in an intermediate course in quantum 
mechanics. Our explanation for this curious situation is that, originally, superselection rules 
were introduced in the context of quantum field theory and they were considered features of 
infinite dimensional theories. On the other hand, perhaps some  theoretical prejudices make 
people believe that there was impossible to find superselection rules in simple quantum systems.  
However, we now know that one can find them even in rather large class of one-dimensional  
quantum systems,  those precisely that have received  attention because, it was believed, could 
furnish counterexamples to the non-degeneracy theorem for bound states in one-dimensional quantum 
systems; a notable example of this class of systems is the one-dimensional hydrogen atom 
(N\'u\~nez-Y\'epez \ea\ 1987). \Par

It is also interesting to point out that the so-called enviromentally induced superselection 
rules (Zureck 1982) could perhaps give a way out to one of the main interpretation problems of 
quantum mechanics, simbolized by the Schr\"odinger cat paradox. The point is, if it could be 
established the existence of enviromentally induced superselection rules between certain different
macroscopic states this would forbid at once the spooky possibility of superposing dead  with 
alive cat states. \Par
					 
\noindent{\bf Acknowledgements}\Par

\noindent  This work has been partially supported by CONACyT (grant 1343P-E9607). ALSB wants to 
thank J L Carrillo-Estrada and O L Fuchs-G\'omez for their one-week hospitality at, respectively, 
the Instituto de F\'{\i}sica and the Facultad de Ciencias F\'{\i}sicas y Matem\'aticas of  BUAP, 
where this work was partially written. We are also extremely grateful to P Pfeifer for providing 
us with a copy of his dissertation. The useful comments of P A Terek, M Tlahui, Ch Mec, G Billi, 
M Miztli, Ch Dochi, F.\ Sadi,  B.\ Caro, and F C Bonito are gratefully acknowledged. Last but not least, HNNY and ALSB  want to  dedicate this work to  B Minina, M Mina, B Kot, Q Motita and K  Quiti. \Par

\vfil
\eject

\noindent{\bf References} \Par
\baselineskip 14 pt

\noindent Aharonov Y and Susskind L 1967 Phys.\ Rev.\ {\bf 155} 1428 \Par

\noindent Bargman V 1954 Ann.\ Math.\ {\bf 59} 1\Par

\noindent Boya L J, Kmiecik M,  and Bohm A  1988 Phys.\ Rev.\ A {\bf 37} 3567 \Par

\noindent Cooper F, Khare A and Sukhatme I 1995 Phys.\ Rep.\ {\bf 251} 267 \Par

\noindent Feynman R P, Leighton R B, and Sands M 1965 {\it The Feynman 
Lectures on Physics} Vol 3 (Reading: Addison-Wesley) Ch 3 and 4\Par

\noindent Foldy L L 1954 Phys.\ Rev.\ {\bf 93} 1395 \Par

\noindent Feynman R P and Weinberg S 1987 {\it Elementary particles and the laws of physics. The 1986 Dirac memorial lectures} (Cambridge: Cambridge University Press) \Par

\noindent Galindo A, Morales A and N\'u\~nez-Lagos R 1962 J.\ Math.\ Phys.\ {\bf 3} 324 \Par

\noindent Hegerfeld G C, Kraus K, and Wigner E P 1968  Phys.\ Rev.\ {\bf 9} 2029 \Par

\noindent Kaempfer F A 1965 {\it Concepts in quantum mechanics} (New York: Academic) Appendix 7\Par

\noindent Landau L and Lifsitz E M 1977 Quantum Mechanics (Oxford: Pergamon) \S 14
\Par

\noindent Lipkin H J 1973 {\it Quantum mechanics: new approaches to selected topics} (Amsterdam: North Holland) Ch 5 \Par
  
\noindent Mart\'{\i}nez-y-Romero R P, N\'u\~nez-Y\'epez H N, Vargas C A and Salas-Brito A L 1989a Rev.\ Mex.\ Fis.\ {\bf 35} 617 \Par

\noindent Mart\'{\i}nez-y-Romero R P, N\'u\~nez-Y\'epez H N  and Salas-Brito A L 1989b Phys.\ Lett.\ A {\bf 142} 318 \Par

\noindent M\"uller-Herold U 1985 J.\ Chem.\ Educ.\ {\bf 62} 379 \Par

\noindent N\'u\~nez-Y\'epez H N, Vargas C A and Salas-Brito A L 1988 J.\ 
Phys.\ A: Math.\ Gen.\  {\bf 21} L651\Par

\noindent N\'u\~nez-Y\'epez H N and Salas-Brito A L 1987 Eur.\ J.\ Phys.\ {\bf 8} 307 \Par

\noindent N\'u\~nez-Y\'epez H N, Vargas C A and Salas-Brito A L 1987 Eur.\ J.\ Phys.\ {\bf 8} 189 \Par

\noindent N\'u\~nez-Y\'epez H N, Vargas C A and Salas-Brito A L 1989 Phys.\ Rev.\ A {\bf 39} 4306 \Par

\noindent Pfeifer P 1980 {\it Chiral molecules: A superselection rule induced by the radiation field} ETH Dissertation \Par

\noindent Pfeifer P 1981 in  Gustaf\-son K E and Reinhardt P Eds  {\it Quantum Mechanics in mathematics, chemistry, and physics} (New York: Plenum) p 315 \Par

\noindent Pfeifer P 1983 in Hinze J Ed {\it Energy storage and redistribution in molecules} (New York: Plenum) p 315 \Par

\noindent Primas H 1981 {\it Chemistry, quantum mechanics and reductionism} Lecture Notes in Chemistry Vol 24 (Berlin: Springer)\Par

\noindent Putterman S 1983 Phys.\ Lett.\ A {\bf 98} 324 \Par

\noindent Robinett R W 1997 {\it Quantum Mechanics} (Oxford: Oxford University Press) \S 14.4 \Par

\noindent Salas-Brito A L 1990   {\it \'Atomo de hidr\'ogeno en 
un campo magn\'etico infinito: Un modelo con regla de superselecci\'on}  Ph D Thesis FC-UNAM \Par

\noindent Salas-Brito A L, Mart\'{\i}nez-y-Romero R P, N\'u\~nez-Y\'epez H N and Vargas C A 1990 
in Garibotti C Ed {\it Actas del 3er Encuentro Latinoamericano sobre 
Colisiones At\'omicas, Moleculares y Electr\'onicas}, (Bariloche: CNEA) p 102 \Par

\noindent Sakurai J J 1985 {\it Modern quantum mechanics} (Reading: Addison-Wesley) \S 3.4 \Par

\noindent Shilov  G E 1977 {\it Linear algebra} (New York: Dover) p 45 \Par

\noindent Streater R F and Wightman A S 1964 {\it PCT, spin and statistics, and all that} (New 
York:  Benjamin) Ch 1 \Par

\noindent Strocchi F and Wightman A S 1974 J.\ Math.\ Phys.\ {\bf 15} 2198 \Par

\noindent Sudbery A 1986 {\it  Quantum mechanics and the particles of 
nature} (Cambridge: Cambridge University Press) Ch 5 \Par

\noindent  Weinberg S 1995 {\it The quantum theory of fields} Vol 1 (Cambridge: Cambridge 
University Press) p 90 \Par 

\noindent Wick G C, Wightman A S and Wigner E 1952 Phys.\ Rev.\ {\bf 88} 
101 \Par

\noindent Zureck W H 1982 Phys.\ Rev.\ D {\bf 26} 1862 \Par

 \vfill
 \eject
  \end